\documentclass[a4paper,UKenglish]{oasics-v2018}

\usepackage{microtype}

\title{Identifying Barriers to Adoption for Rust through Online Discourse}

\author{Anna Zeng}{Stanford University}{}{}{}

\author{Will Crichton}{Stanford University}{}{}{}

\authorrunning{A. Zeng and W. Crichton}

\Copyright{Anna Zeng and Will Crichton}

\subjclass{Human-centered computing $\rightarrow$ Human computer interaction (HCI)}


\category{}

\relatedversion{}

\supplement{}

\funding{}

\acknowledgements{}

\EventEditors{Titus Barik, Joshua Sunshine, and Sarah Chasins}
\EventNoEds{3}
\EventLongTitle{9th Workshop on Evaluation and Usability of Programming
Languages and Tools (PLATEAU 2018)}
\EventShortTitle{PLATEAU 2018}
\EventAcronym{PLATEAU}
\EventYear{2018}
\EventDate{November 5, 2018}
\EventLocation{Boston, Massachusetts, USA}
\EventLogo{}
\SeriesVolume{67}
\ArticleNo{5} 
\nolinenumbers
\hideOASIcs  

\newcommand{\hypothesis}[2]{\noindent\textbf{Hypothesis #1}: \textit{#2}}
\usepackage{comment}

\begin{document}

\maketitle

\begin{abstract}
Rust is a low-level programming language known for its unique approach to memory-safe systems programming and for its steep learning curve. To understand what makes Rust difficult to adopt, we surveyed the top Reddit and Hacker News posts and comments about Rust; from these online discussions, we identified three hypotheses about Rust's barriers to adoption. We found that certain key features, idioms, and integration patterns were not easily accessible to new users.
\end{abstract}

\section{Introduction}

Rust is a new
programming language designed to usher low-level programming into the modern era. Rust uses strong type systems and functional programming to execute programs efficiently
while avoiding the many safety problems that plague C and C++. As an open-source project with support from Mozilla, the Rust ecosystem has grown rapidly over the last decade. Hundreds of companies deploy Rust in production, and thousands of developers regularly use Rust in their projects.
However, Rust has a notoriously steep learning curve.
A community survey in 2017 revealed that 25\% of the people who tried Rust and dropped it felt the language was ``too intimidating, too hard to learn, or too complicated'' \cite{rustsurvey2017}.
Prominent members of the Rust community have stated that Rust is \textit{supposed} to be hard to learn: ``Rust has never claimed that it is something you can learn in half a week'' \cite{hardtolearn}. 
Even still, the learning curve is only one aspect of many (library support, tooling, compile times, etc. \cite{meyerovich2013empirical}) that influences the adoption of a programming language.

If Rust is hard to learn or use, the question becomes: how should its developers prioritize language features to drive adoption, addressing the key challenges facing its current or potential user base? A developer on the Rust compiler told us in an interview that these decisions are usually made in an ad hoc way, based on the intuitions of the language developers and occasional feedback from community members in a potpourri of online forums \cite{niko_interview}. To address this issue, we sought to understand the challenges of adopting Rust by analyzing online discourse within the Rust community. We conducted a holistic survey of popular posts and comment threads about Rust to identify beliefs of developers using Rust that highlight ongoing issues in the practice of the language.


\section{Methodology}

Prior work on analyzing language adoption has focused on evaluating various languages and tools through questionnaires, surveys, in-person interviews, and automated code base analysis. For example, Meyerovich and Rabkin \cite{meyerovich2013empirical} combined all of the above to highlight both empirical popularity trends as well as reported beliefs of developers about a wide range of topics from types vs. tests to features vs. libraries. Christakis and Bird \cite{christakis2016developers} used surveys and interviews to understand the adoption of program analyzers, and Ray et al. \cite{ray2014large} used program analyzers to compare languages based on bugs detected in a large corpus of GitHub repositories.

This study differs from previous work 
in two ways: First, we focused only on a single language, Rust. While this focus potentially limits the generality of our insights, it allows us to produce deeper insights via thorough consideration of the factors driving specifically Rust's adoption. Second, we performed a content analysis on the existing discourse on Rust rather than creating a new survey. We observed that Rust is frequently written about on blogs, and subsequently discussed online by the people developing Rust itself (henceforth referred to as ``Rust language developers''), by the people learning and using Rust (``Rust community''), and by the broader tech community. In fact, a version of this ethnomethodological approach \cite{garfinkel1967studies} was already used by the Rust language developers to inform the Rust 2018 roadmap \cite{rustroadmap2018} and also used to analyze other online communities like Mechanical Turk \cite{martin2014being}.

We gathered a corpus of articles and corresponding comments from Hacker News (HN), a forum for general tech-centric discussions, and the /r/rust subreddit, a forum specifically about Rust.  We selected these two forums both because they are frequent hosts to discussions about Rust and use upvote mechanisms to sort content. Upvotes act as a loose proxy for what readers consider to be good contributions to the conversation or perspectives they agree with, and have been shown to correlate with coarse notions of quality \cite{stoddard2015popularity}. While upvote-based filtering can reduce exposure to controversial opinions, given the vast amount of possible content to read, we found it critical in improving the signal-to-noise ratio. To select the final corpus, we filtered for HN articles with ``Rust'' in the title, and considered all /r/rust articles. Then, we filtered for posts with at least 250/200 upvotes on HN/Reddit respectively, a total of 424 posts. From there, we selected posts that we felt were most relevant towards understanding user experiences in Rust, e.g. choosing ``Three months of Rust'' and ``Why I'm dropping Rust' over ``Announcing Rust 1.12''.  This ad hoc filter was not applied soundly or completely (we read as much as time permitted), so we do not claim our survey is exhaustive. Our final corpus contains 50 posts, with corresponding comment sections on both forums where applicable.

For each article, we performed a content analysis on both the document and at least the top five comments by upvotes for each forum the article appeared on. 
Then we categorized the articles (categories like ``community'', ``ergonomics'', ``tooling'', ``security'', etc.) and looked for trends within each category, forming preliminary hypotheses about barriers to adoption that were both novel and actionable to the Rust community.
From the insights, we formed hypotheses that help explain the experiences that users encounter in our corpus.
Concurrently, we interviewed three Rust language developers to help us contextualize our findings and to understand prevailing attitudes towards Rust's usability in the community.

\section{Hypotheses}

\hypothesis{1}{Rust is primarily promoted for safety and speed; while those aspects matter to users, the tooling around Rust is equally valuable, but its value is not as clearly communicated by the language developers.}

Potential users need to understand Rust's features and goals in order to determine whether to use it. The authoritative source of information on Rust is its official website, \url{https://rust-lang.org}. On the front page and the FAQ, Rust is promoted as ``a systems programming language that runs blazingly fast, prevents segfaults, and guarantees thread safety.'' Laundry lists of language features including ``zero-cost abstractions,'' ``pattern matching,'' and ``type inference'' are also provided. The standard library, the tooling, and the ecosystem are all absent from this messaging \cite{rust_homepage}.

To evaluate whether Rust users found the language useful for its claimed benefits, we analyzed 12 experience reports (e.g.  ``Trying out Rust for Graphics Programming'') and 6 language comparisons (e.g. ``Comparing Rust and Java''). Across the 18 articles, we counted the reported pros/cons of Rust. The first and third most reported benefits of Rust were elimination of runtime errors (7 articles) and data races (5 articles). Runtime errors include both avoiding memory errors through Rust's memory safety analyzer as well as avoiding unhandled failures through sum types, e.g. \verb|Option<T>|. This finding is consistent with Rust's messaging---users empirically self-report the utility of Rust's safety guarantees.

By contrast, the second most reported benefit of Rust was Cargo, Rust's build system and dependency manager (6 articles). One article summarized the collective sentiment \cite{rustthings}:

\begin{quote}
Instead of having to invoke \verb|pkg-config| by hand or with Autotools macros, wrangling include paths for header files and library files, and basically depending on the user to ensure that the correct versions of libraries are installed, you write a \verb|Cargo.toml| file which lists the names and versions of your dependencies. [...] It just works when you \verb|cargo build|.
\end{quote}

\noindent The Rust language developers likely understand the importance of Rust's tooling, being a major part of the Rust 2017/18 roadmaps. However, because Rust's promotional messaging doesn't clearly emphasize these features, this absence suggests a possible disconnect between what the language developers and potential users consider the most important features of Rust.

\

\hypothesis{2}{Complex pointer aliasing patterns are primarily implemented through existing libraries built on unsafe code, but Rust users have a hard time discovering these solutions.}

To guarantee memory safety for low-level programs with direct access to memory, Rust employs the ``borrow checker'', a static analysis tool that prevents memory-unsafe operations, e.g.\ returning a dangling pointer to a stack-allocated value. The borrow checker does not permit mutable aliases, or holding two mutable pointers to the same piece of memory. Aliasing patterns that violate this rule, like reference-counted pointers, can be implemented carefully through Rust's \verb|unsafe| construct, often provided as libraries. However, this restriction is often daunting to Rust novices
because it disallows patterns that are easy to express in other low-level languages, and it makes solutions difficult to find. Out of the 18 experience reports and language comparisons, the complexity of the borrow checker was the second most frequently mentioned complaint (only behind compiler version issues of stable vs. nightly). For example, one user implementing a video codec found \cite{codec}:

\begin{quote}
Video codecs usually operate on planes and there you’d like to operate with different chunks of the frame buffer (or plane) at the same time. Rust does not allow you to mutably borrow parts of the same array even when it should be completely safe like \verb|let mut a = &mut arr[0..pivot]; let mut b = &mut arr[pivot..];|.
\end{quote}

\noindent Another user implementing a GUI framework found \cite{droppingrust}:

\begin{quote}
For \verb|nanogui|... each widget has a pointer to a parent and a vector of pointers to its children. How does this concept map to Rust? There are several answers:
  1. Use a naive \verb|Vec<T>| implementation.
  2. Use \verb|Vec<*mut T>|.
  3. Use \verb|Vec<Rc<RefCell<T>>>|.
  4. Use C bindings.
... I have tried options 1 through 3 with several drawbacks, each making them not fit for use. I'm currently looking at point 4 as my only remaining option to use.
\end{quote}

\noindent In both cases, Rust users encountered a particular memory access pattern (disjoint mutable pointers to an array, widget trees with back pointers) that Rust disallowed. As with other complex aliasing patterns like reference counting, the solution to these types of problems is sufficiently complex that Rust users aren't expected to implement it themselves, but instead defer to external code. In both cases above, commenters pointed out standard library functions (\verb|Vec::split_at_mut|) and third-party libraries (\verb|petgraph|) for solving these issues respectively; however, the authors were not able to independently discover these solutions. These experiences suggest that Rust needs better resources to help users identify common aliasing patterns and understand what tools exist to solve those problems.

\

\hypothesis{3}{Although incremental migration of existing codebases into Rust seems promising, Rust users aren't pursuing this path because the cost of integrating Rust into a different language ecosystem or toolchain is too great.}

One important path to adoption of Rust is incremental migration, or gradually rewriting components of a large software system from a host language (like C/C++) into Rust. Mozilla's initial motivation for Rust was to replace performance-critical parts of Firefox (Project Servo), and others have begun exploring integrating with databases (Postgres) and operating systems (Linux). However, the idea of incremental migration is almost entirely absent from the discourse we studied. Only one of our 50 articles directly described experiences integrating Rust into an existing system; most experience reports detailed new projects in Rust or entire rewrites of existing projects. While lack of discourse doesn't necessarily imply lack of adoption, it 
still does not bode well for the path of incremental migration---issues out of the public eye are unlikely to receive attention either from the Rust compiler developers in prioritizing work, or from the Rust community in generating documentation and tooling.

A possible explanation 
is that the challenges of incremental migration of Rust may be surmountable for larger teams, like those supporting Firefox, but are still too challenging for most Rust users. Discussion of incremental migration in our corpus largely occurred in scattered comments focused on describing challenges, not hailing great successes. For example, a Servo developer described in an interview the challenges of bridging Rust, C++, and Javascript in Firefox: the ease of accidentally invalidating a C++ reference when in Rust, the challenge of managing macros across the language barriers, and the complexity of tracking when each variable would be deallocated by which runtime \cite{jdm_interview}.

Another example of a difficult challenge for most Rust users is working on Rust without Cargo, a tool which reduces the overhead of starting (and maintaining) a new project and reusing third-party Rust code. In low-level programming, build systems and package management are traditionally relegated to a hodgepodge of tools like Make, Autotools, CMake, and Apt; in contrast, the Rust community uses Cargo for both build process and dependency management. While Cargo is not strictly required, all major libraries must be built with Cargo. Developing Rust without Cargo means losing easy access to these libraries.
For example, Facebook's Mononoke project, a rewrite of Mercurial in Rust, initially could not use any packages outside the standard library due to integration requirements with Facebook's custom build system, which significantly slowed adoption of external libraries. The lack of discourse around these issues suggest more resources should potentially be dedicated to reducing these barriers to adoption for incremental migration into Rust.

\section{Discussion}

By applying an ethnomethodological approach to Rust's online discourse, we identified supporting evidence for three hypotheses about factors which meaningfully influence the adoption of Rust. This methodology is neither fully precise nor conclusive, but it provides a useful signal to direct further evaluation of these hypotheses. While online commenters are not subject to the same level of rigor as peer-reviewed research, our experience suggests that blog posts and forum threads still contain an enormous amount of collective wisdom that is perhaps under-appreciated in academic literature. 
Much of the prior work has trended towards precision through controlled experiments or
breadth through surveys and code base analyses.
However, we believe that understanding these online communities can provide valuable guidance to PL/HCI researchers seeking to address the key problems facing today's programmers.

To that end, one key question is this study's replicability: could other researchers recreate our results for Rust, or perform the same study on other programming languages? Due to the time-consuming nature of manual content analysis, consistently filtering articles from the larger corpus is important. Creating better filters would be simplified by looking at a smaller domain of documents with more quantifiable filters, e.g. analyzing just experience reports, or articles about security, or articles about language design decisions. Conversely, studies that attempt to just categorize the discourse by identifying common topics can reduce work done in later studies on any given category. Lastly, applying a consistent content analysis methodology is challenging given the open-ended, free-form nature of synthesizing connections in document surveys. Perhaps here the PL/HCI community could provide more guidance on standard methodologies for ethnomethodological analyses.

The second question is the applicability of this study's results: what further experiments do our results suggest? We believe the logical next step is to develop surveys targeted towards understanding the extent of the issues identified above. While usability surveys often resort to generic questions like ``where do you think Rust can improve?'' (e.g. as in \cite{rustsurvey2017}), the hypotheses above suggest more specific questions like:
\begin{itemize}
\item How has your perception of Cargo's value changed between starting with Rust and today?
\item What parts of Rust do you find most useful that you weren't told about initially?
\item What are examples of times when you couldn't solve an issue with the borrow checker after several hours? If you eventually solved it, how did you solve it? 
\item What factors influence your likelihood to adopt Rust into an existing non-Rust project?
\end{itemize}
\noindent Given feedback on these questions from the community, future work could explore controlled experiments for more replicable identification of usability issues, as well as analyze how insights gleaned from Rust can generalize to other programming language communities.

\bibliography{main}

\end{document}